\documentclass{aastex63}

\usepackage{amsmath}

\begin{document}

\title{A FIRST SPECTROSCOPIC MEASUREMENT OF THE MAGNETIC FIELD STRENGTH FOR AN ACTIVE REGION OF THE SOLAR CORONA}

\correspondingauthor{Tomas Brage}
\email{tomas.brage@fysik.lu.se}
\correspondingauthor{Roger Hutton}
\email{rhutton@fudan.edu.cn}

\author{Ran Si}
\affiliation{Division of Mathematical Physics, Department of Physics, Lund University, 221 00 Lund, Sweden}
\affiliation{Spectroscopy, Quantum Chemistry and Atmospheric Remote Sensing (SQUARES), CP160/09, Universit\'{e} libre de Bruxelles, Av. F.D. Roosevelt 50, 1050 Brussels, Belgium}
\author{Tomas Brage}
\affiliation{Division of Mathematical Physics, Department of Physics, Lund University, 221 00 Lund, Sweden}
\author{Wenxian Li}
\affiliation{Division of Mathematical Physics, Department of Physics, Lund University, 221 00 Lund, Sweden}
\affiliation{Department of Materials Science and Applied Mathematics, Malm\"{o} University, SE-20506 Malm\"{o}, Sweden}
\author{Jon Grumer}
\affiliation{Theoretical Astrophysics, Department of Physics and Astronomy, Uppsala University, Box 516, SE-751 20 Uppsala, Sweden}
\author{Meichun Li}
\affiliation{Shanghai EBIT Laboratory, Key Laboratory of Nuclear Physics and Ion-beam Application, Institute of Modern Physics, Department of Nuclear Science and Technology, Fudan University, 200433 Shanghai, China}
\affiliation{School of Electronic Information and Electrical Engineering, Huizhou University, 516007 Huizhou, China}
\author{Roger Hutton}
\affiliation{Shanghai EBIT Laboratory, Key Laboratory of Nuclear Physics and Ion-beam Application, Institute of Modern Physics, Department of Nuclear Science and Technology, Fudan University, 200433 Shanghai, China}

\begin{abstract}
For all involved in astronomy, the importance of monitoring and determining astrophysical magnetic field strengths is clear. It is also a well-known fact that the corona magnetic fields play an important part in the origin of solar flares and the variations of space weather. However, after many years of solar corona studies, there is still no direct and continuous way to measure and monitor the solar magnetic field strength. We will here present a scheme which allows such a measurement, based on a careful study of an exotic class of atomic transitions known as magnetic induced transitions in Fe$^{9+}$. In this contribution we present a first application of this methodology and determine a value of the coronal field strength using the spectroscopic data from HINODE.
\end{abstract}

\section{Introduction} \label{sec:intro}
Magnetic field holds a central position within solar research, continuous or on-demand measurements of the magnetic fields in the solar corona remain one of the major challenges in solar physics~\citep{Casini2017}. It is of importance to the prediction of solar events such as flares or coronal mass ejections, and ultimately for space-weather forecasting to avert damage to navigation and communication satellites, interference with airplane
navigation systems and disruptions in power grids which could cause large-scale black-outs~\citep{Schrijver2015}.
Due to the potential threat to society and human well-being from variations in the space weather, it is important to develop methods to continuously monitor the magnetic fields of the corona and measure their strengths.
The very recent inauguration of NSF’s DKIST~\citep{Tritschler2016} and the launchings of the NASA mission Parker Solar Probe (PSP, launched in August 2018) and the ESA/NASA mission Solar Orbiter (launch in February 2020) observatories
are forming an unprecedented solar corona and inner heliospheric campaign targeted at understanding how
stars create and control their magnetic environments~\citep{MartinezPillet2020}.
Unfortunately, a candidate for such a measurement has eluded the solar physics community. This may be the largest single factor blocking progress in coronal physics and is hindering the attempts to answer questions related to coronal heating, the triggering of flares and coronal mass ejections, as well as the acceleration of the fast and slow solar wind~\citep{Solanki2006}.

To address this, we have over the past few years been investigating a spectroscopic method based on quantum-interference effects in the Fe$^{9+}$ ion~\citep{Li2015,Li2016,Judge2016,Si2020}. This particular interference is caused by magnetic fields external to the ion, and hence this idea has the potential to act as a probe of the coronal field strengths. In this contribution we present a first application of this methodology and determine the values of the coronal field strengths using pre-existing spectroscopic data from  Extreme-Ultraviolet Imaging Spectrometer (EIS) on the
Hinode satellite~\citep{Culhane2007,Brown2008}. The measurement is even fast enough, relative to the lifetime of solar flares, that we could track the development of the field as the flare develops.

Observing a single spectral line from an element in a certain charge state provides little information about the environment in which it was emitted, while a group of lines can give us a much more detailed picture.
Just to give an example on how two lines can be used, their intensity ratio - especially if close-by in wavelength and from a single atomic charge state - often acts as a probe of the local electron density and temperature in the plasma~\citep{Feldman1978}. To determine local plasma properties, the strategy is therefore reduced to finding a pair of lines of similar wavelength, where one originates from an upper level with a radiative decay rate of the same order of magnitude as the electron collisional de-excitation rate. The other line, which we will refer to as the normalization line, should have a radiative decay rate that is signifcantly faster and therefore its decay rate is insensitive to the collisional rates.

The same principle, where the intensity of one spectral line is sensitive and one insensitive to a certain environmental variable, can be used to measure other plasma properties. In this letter we will discuss the strength of the magnetic field local to the observed ions and how it can induce new lines, so called magnetic-field induced transitions (MIT's). It has been illustrated that the intensity of these lines can show a strong, to first order quadratic dependence on the external field strength~\citep{Grumer2014}. Since the magnetic fields inside the ions are enormous (in the ions of interest here on the order of hundreds or even thousands of Tesla), it requires, in the general case, strong external fields to induce these lines from perturbations of the atomic structure. As an example, in Ne-like ions and for field strengths of a few Tesla, MIT's have been observed using an Electron Beam Ion Trap ~\citep{Beiersdorfer2003,Beiersdorfer2016}.
Such cases are not of particular interest in solar physics, since the strongest fields we can observe are in sunspot, and they are always weaker than 1 T.

The rate, and therefore the intensity, of an MIT of electric dipole (E1) type is to first order given by
\begin{equation}\label{eq:Amit}
    A_\mathrm{MIT} \propto A_\mathrm{E1}\frac{B^2}{\left(\Delta E\right)^2}
\end{equation}
where $B$ is the strength of the external magnetic field, $\Delta E$ is the energy separation between
a metastable level and a close-by upper level of a fast E1 transition, in the following referred to as the feeding level, while $A_\mathrm{E1}$ is the decay rate of the feeding level.
The metastable and feeding levels are mixed in the presence of an external magnetic field, which causes them to share their properties. This leads to the emergence of a new radiative transition from the metastable level, the MIT. Therefore, if one can find a metastable level which is close to a short-lived one in an abundant atomic charge state, one can expect that MIT's could be observed in the presence of a magnetic field. Of particular interest are those cases where this energy separation is small enough to cause a pseudo-degeneracy, which leads to a dramatic increase in the MIT rate of Eq. \ref{eq:Amit} and thus also in the sensitivity of the rate to the magnetic field. Such degeneracies are however not necessarily predicted by the symmetries and the gross structure model of the ion, but could instead occur by chance as a result of a rather complex atomic structure.

A search for an atomic system involving a suitable pseduo-degeneracy was initiated a few years ago~\citep{Li2015,Li2016,Judge2016,Si2020}, motivated by its potential as a magnetic-field probe. This led to the discovery of Fe$^{9+}$ where the excited levels $3p^43d\ {^4}\mathrm{D}_{5/2}$ and ${^4}\mathrm{D}_{7/2}$ fulfill the requirements of being very
close in energy and having significantly different lifetimes. Fortunately, for astrophysical applications, this pseudo-degeneracy exists in an iron-ion with a large abundance in many celestial objects, amongst them the Sun.
It was found that this system could give rise to a considerable MIT even for field strengths of the same order of magnitude as one might expect in the solar corona, which presently are inaccessible from direct measurements.
This motivated further studies of the Fe$^{9+}$ system and investigations of its potential as a probe of coronal magnetic fields. However, the pseudo-degeneracy of ${^4}\mathrm{D}_{5/2,7/2}$ makes the two lines to the ground state a challenge to resolve  since the transitions are in the VUV-region whereas the energy difference is only around 3.6 cm$^{-1}$, to move to an actual determination of the coronal field this method had to be refined and supported by complex atomic and solar models. In this report we can finally, for the first time, present a direct determination of these field strengths.

\section{MODELLING AND ANALYSIS}
For the spectral modelling we use the \textsc{ChiantiPy} spectral synthesis code,  which is tailored for interpretation of spectra from high-temperature, optically-thin astrophysical sources,
together with electron collision data from the Chianti database~\citep{DereK.P.1997,Landi2013} and radiative transition data from~\citet{Wang2020}.

\begin{figure*}
    \centering
    \includegraphics[width=0.5\textwidth]{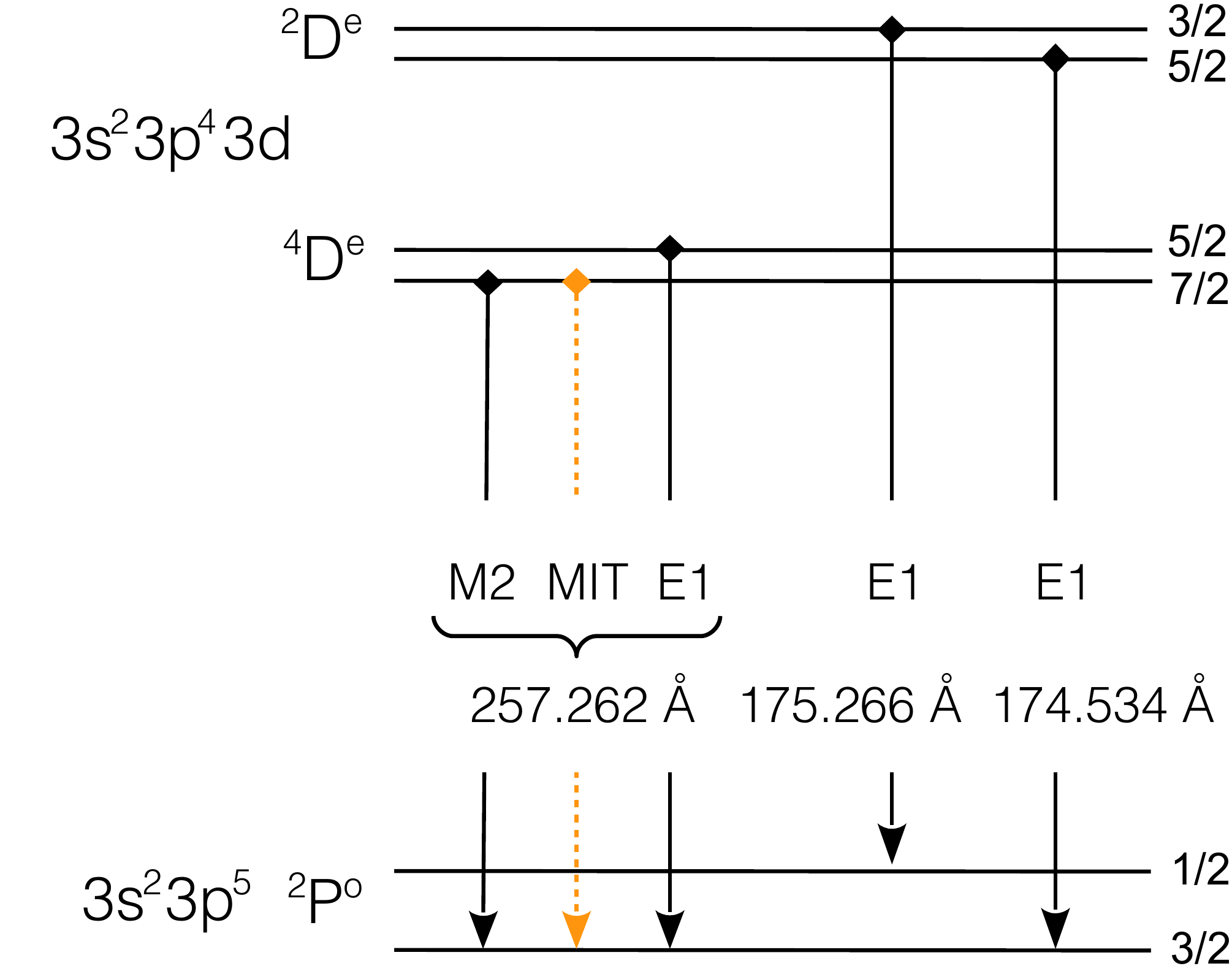}
    \caption{\label{fig_lev}Schematic energy level diagram and decay channels for the levels in Fe$^{9+}$ that are relevant to the method discussed in this work (see text). Wavelengths are given in~\AA.}
\end{figure*}

\begin{figure*}
\centering
    \includegraphics[width=9cm]{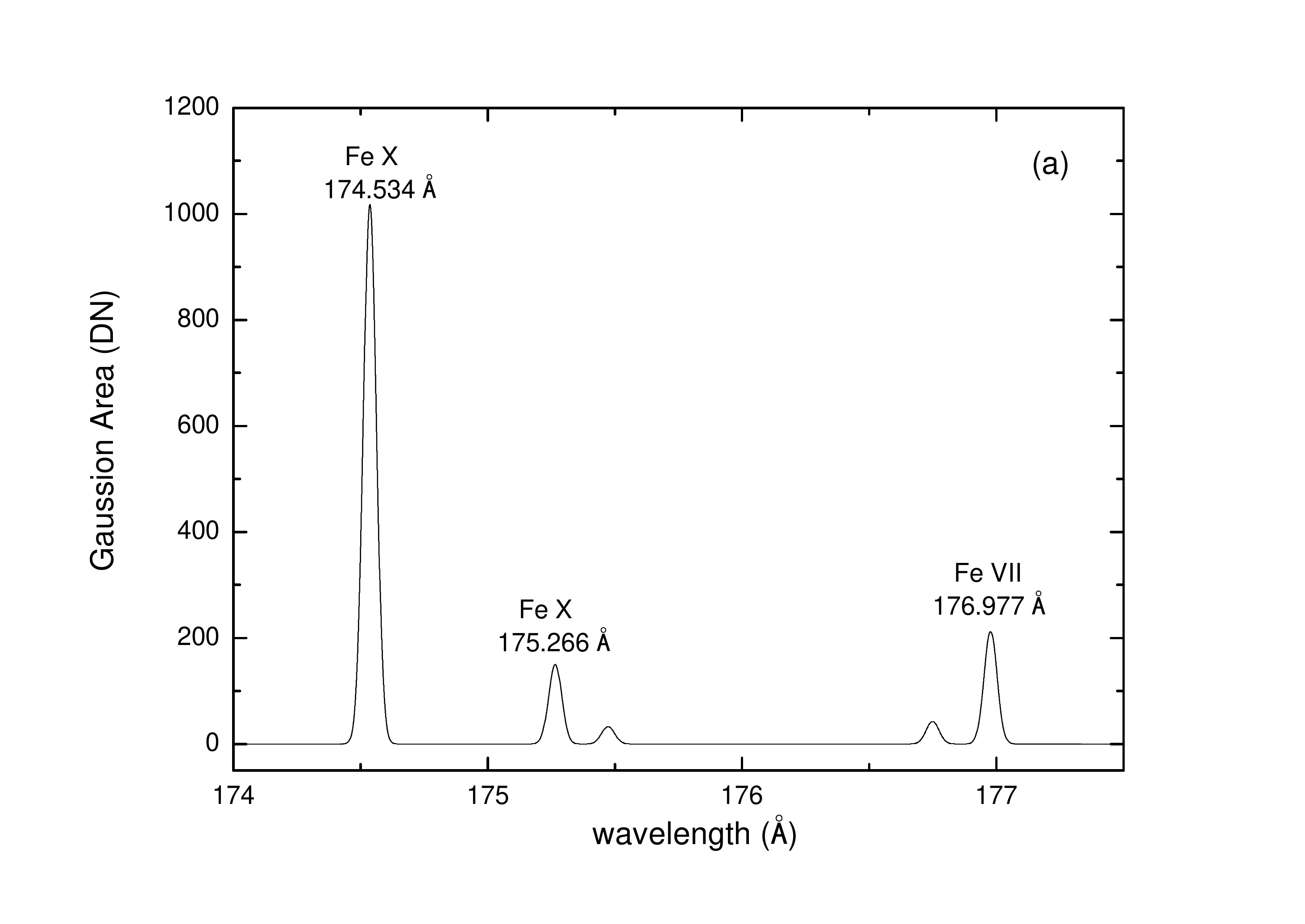}
    \includegraphics[width=9cm]{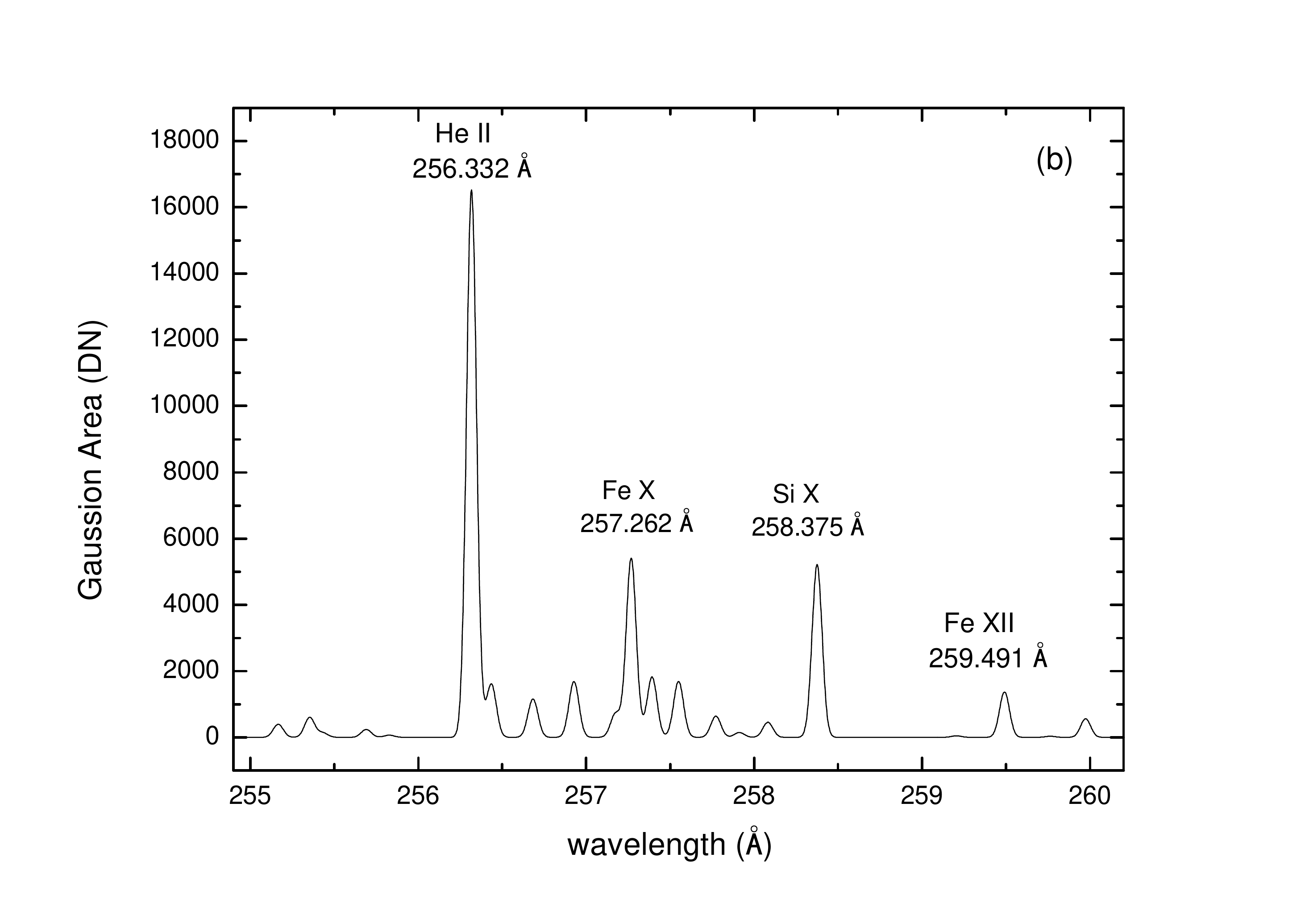}
    \caption{\label{fig_spec} Synthetic Gaussian fitted spectra from the EIS short wave band (a) and long wave band (b) nearby the 257 MIT line (bottom panel) and reference lines (top panel) shown in Fig.~1 for AR2. FWHM$=0.06$~{\AA} for short wave band, FWHM$=0.07$~{\AA} for long wave band, as recommended in~\citet{Brown2008}.}
\end{figure*}

A partial energy level diagram of Fe$^{9+}$ is shown in Fig. \ref{fig_lev}, illustrating the levels and transitions of interest in this work.
The synthetic Gaussian fitted spectra recorded by EIS on Hinode satellite~\citep{Culhane2007,Brown2008} nearby the present interested lines is shown in Fig.~\ref{fig_spec},
demonstrates the lines in Fig.~\ref{fig_lev} are well resolved by EIS.
The main feature for this project is a blended group of three lines marked as having the wavelength of 257.262~{\AA} (denoted by E1, M2 and MIT in Fig.~\ref{fig_lev}).
These are from two different upper levels, one being the $3s^23p^43d\ ^4\mathrm{D}_{5/2}$ decaying to the ground level $3s^23p^5\ ^2\mathrm{P^o_{3/2}}$ through a so-called E1 spin-induced transition.
This level represents the feeding level in our model. The other upper level is the $3s^23p^43d\ ^4\mathrm{D}_{7/2}$ level, with a decay dominated by a forbidden M2 decay, in the absence of external magnetic fields, with a low transition rate. Finally, due to the very near energy degeneracy ($\Delta E\approx 0$) of this level with the $^4\mathrm{D}_{5/2}$ level, an external magnetic field induces a mixing of these two levels causing a magnetic-field induced E1 transition to the $^2\mathrm{P^o_{3/2}}$ ground level.
For the important $\Delta E$ parameter introduced in Eq. \ref{eq:Amit} we use value of $3.6$ cm$^{-1}$
from the most accurate determination to date~\citep{Judge2016}. The two transition rates required to model the MIT spectral line are~\citep{Wang2020},
\begin{align}
  A_\mathrm{E1}({^4}\mathrm{D}_{5/2}\rightarrow {^2}\mathrm{P^o_{3/2}}) & = 6.01\times 10^6~\mathrm{s}^{-1} \; , \\
  A_\mathrm{M2}({^4}\mathrm{D}_{7/2}\rightarrow {^2}\mathrm{P^o_{3/2}}) & = 5.78\times 10^1~\mathrm{s}^{-1} \; .
\end{align}
As pointed out above, in the presence of a magnetic field, the metastable and feeding levels will interact, which is represented by the mixed state
\begin{equation}
|``7/2"\rangle = c_1 |7/2\rangle + c_2|5/2\rangle
\end{equation}
where $J=7/2$ and $5/2$ is the quantum number representing the total angular momentum of the ion, in a field-free space.
The mixing coefficients, $c_1$ and $c_2$ are obtained by determining and diagonalizing the interaction matrix for different magnetic fields, using the \textsc{Grasp} atomic structure suite of programs \citep{ Fischer2019,Fischer2016,Jonsson2013} together with the \textsc{Hfszeeman} add-on module~\citep{Andersson2008, Li2020}.
The $(B/\Delta E)^2$ proportionality of Eq.~\ref{eq:Amit} is due to the $c_2$ mixing coefficient, since an estimated value for the rate in first-order representation can be written as,
\begin{equation}
A_\mathrm{MIT}({^4}\mathrm{D}_{7/2}\rightarrow{^2}\mathrm{P}^o_{3/2}) = \left|c_2\right|^2A_\mathrm{E1}({^4}\mathrm{D}_{5/2}\rightarrow{^2}\mathrm{P}^o_{3/2}) \; .
\end{equation}
By adding this rate for a number of magnetic field strengths to the M2 rate, we can predict the intensity ratio for the combined transition from the ${^4}\mathrm{D}_{7/2,5/2}$ levels, to the normalization lines 
and compare it to the observed ratio from the HINODE data~\citep{Brown2008}.

The decay of the metastable ${^4}\mathrm{D}_{7/2}$ level is dominated by slow M2 and MIT radiative channels resulting in a lifetime of the order of $10^{-2}$~s, implying that the blended spectral feature could be sensitive to the electron densities found in the corona. In order to evaluate these collisional effects, the proposed method thus also requires simultaneous determination of the local electron density.
From the \textsc{ChiantiPy} line intensity modelling at the wavelength range 165~{\AA}$-$290~{\AA} and electron density range $10^8-10^{10}$~cm$^{-3}$, we found that the 174.534~{\AA} line ($3s^23p^43d\ ^2\mathrm{D}_{5/2}-3s^23p^5\ ^2\mathrm{P^o}_{3/2} $ as shown in Fig.~\ref{fig_lev}) is the strongest at all densities, while the relative intensity of the 175.266~{\AA} line   ($3s^23p^43d\ ^2\mathrm{D}_{3/2}-3s^23p^5\ ^2\mathrm{P^o}_{1/2} $) to the 174.534~{\AA} line increases with the electron density.
By adjusting the spectral model to simulate the measured intensity ratio, electron densities of $1.2\times 10^{9}$~cm$^{-3}$ were established for one active region (AR2 in~\citet{Brown2008}).
With these electron densities and the selected normalization lines 174.534~{\AA} and 175.266~{\AA}, magnetic field strengths can finally be determined through comparisons of the spectral model with observed line ratios.

\begin{figure*}
\centering
    \includegraphics[width=9cm]{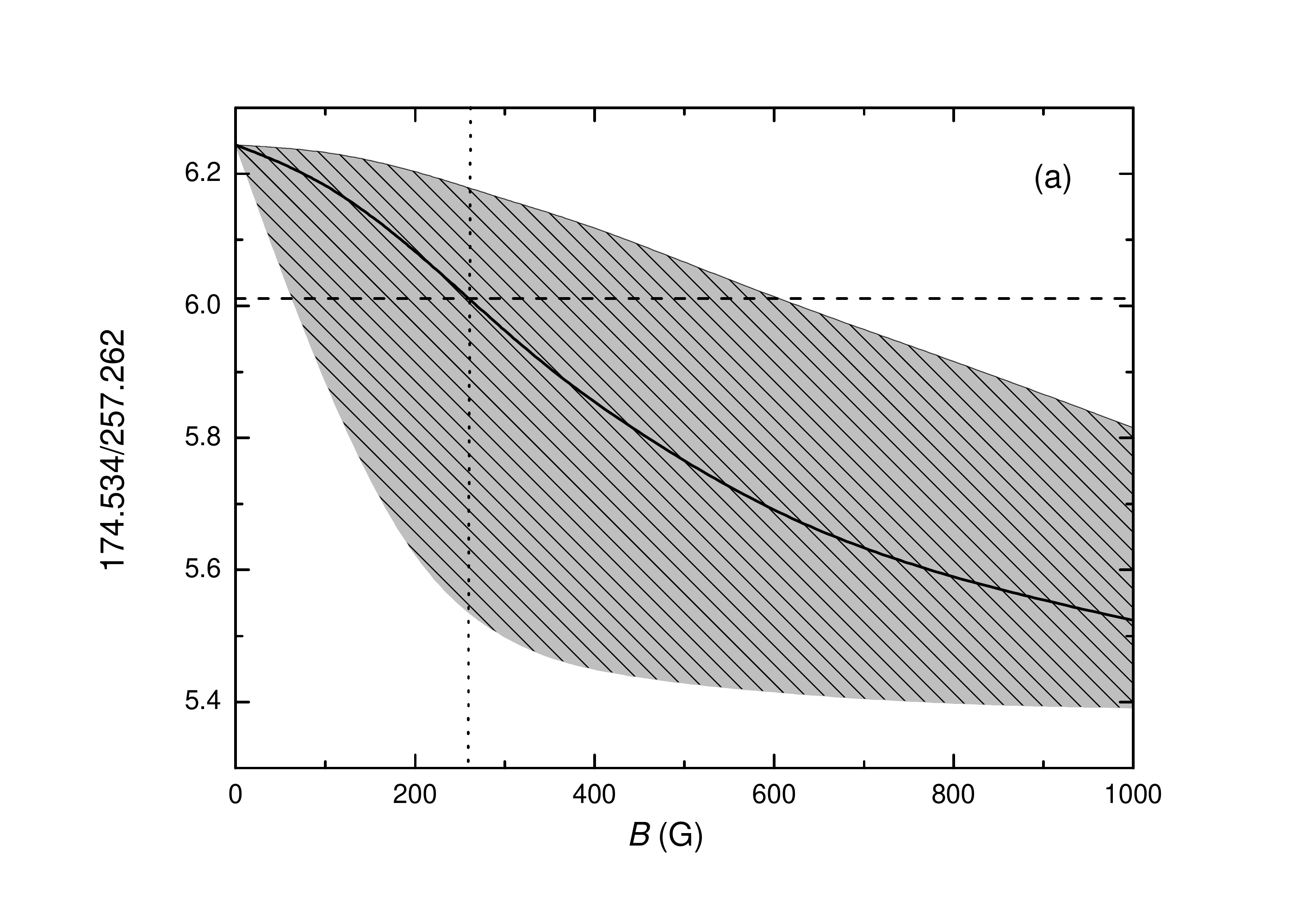}
    \includegraphics[width=9cm]{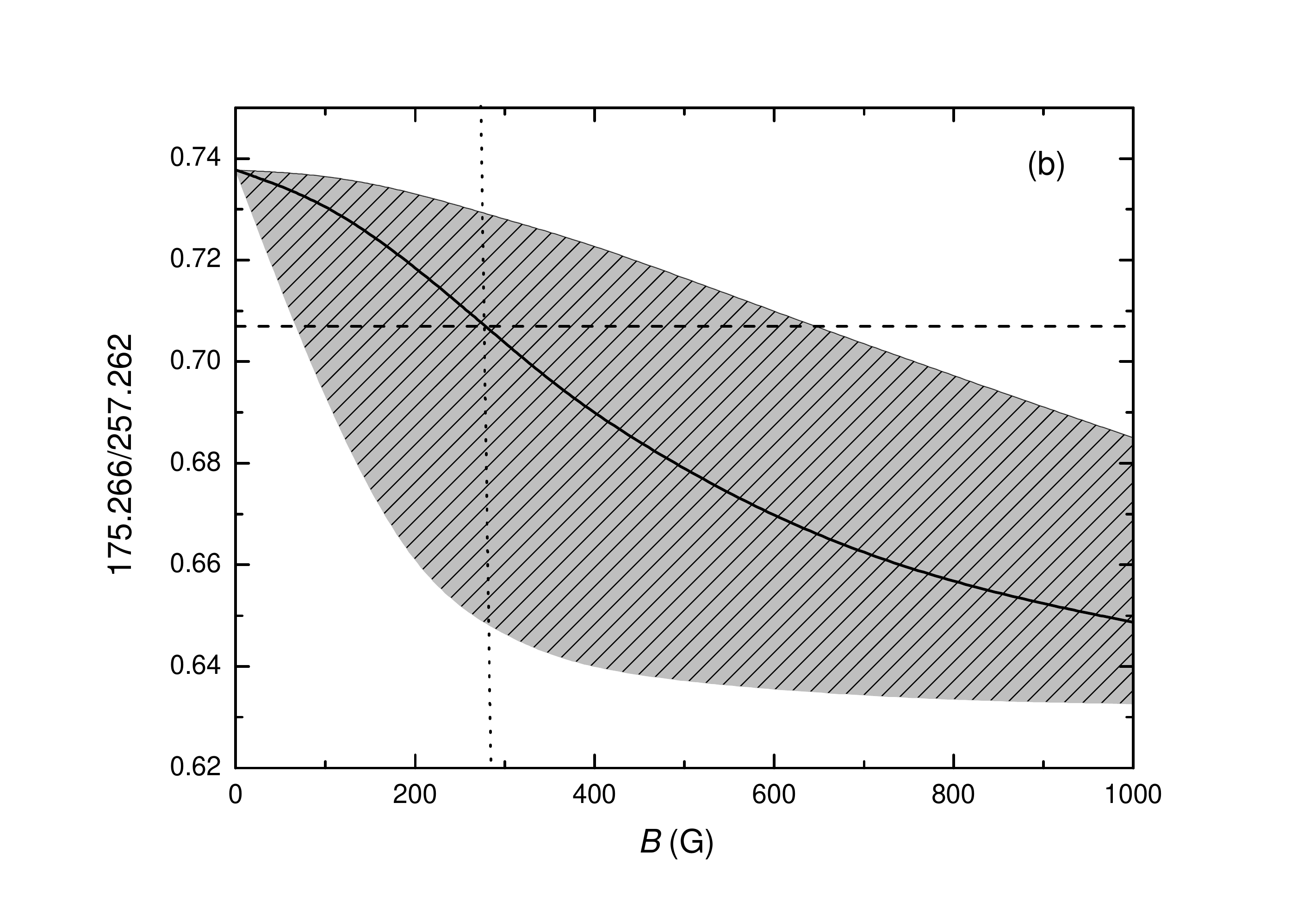}
    \caption{\label{fig_AR2}Simulated intensity ratio (black line) of (a) 174.534/257.262 and (b) 175.266/257.262 as a function of the magnetic field in the AR area. The grey shaded area shows the uncertainty caused by the uncertainty in the $\Delta E$-parameter (see text). The horizontal dashed line is the line ratio measured from HINODE. The vertical dotted line shows our estimated magnetic strength.}
\end{figure*}

By combining theoretical modelling with the observation from HINODE, we estimate solar magnetic fields for AR2.
Fig.~\ref{fig_AR2} present modelled line ratios as functions of magnetic field strengths. Comparing with the observed line intensities from the active region, the best-fit magnetic fields only differ by about four percent, giving an estimated average field of $B_{\mathbf{e}} = 270$~G from $265$~G and $275$~G obtained from the line ratios with lines at 174.534~\AA~(Fig.~\ref{fig_AR2} (a)) and 175.266~~\AA~(Fig.~\ref{fig_AR2} (b)) respectively. This accords with previous estimations of $100-300$~G based on extrapolation from magnetograms at the lower boundary, using a potential or force-free field model~\citep{Aschwanden2014}, however is not a direct measurement of the coronal magnetic fields.

There are a number of uncertainties to be considered, the dominating one coming from the determination of the fine structure energy ($3.6\pm2.7$~cm$^{-1}$~\citep{Judge2016}). The shaded areas of Fig.~\ref{fig_AR2} show limits for the estimated magnetic fields due to the uncertainty in $\Delta E$, with upper and lower boundaries at roughly $B_{\mathbf{e}}/16$ and $3*B_{\mathbf{e}}$, respectively.
There are also possible uncertainties in the atomic data, especially in the M2 and the MIT rates.
The MIT rate depends on the transition rate of the ${^4}\mathrm{D}_{5/2}$ which is in itself a spin-forbidden transition to the ${^2}\mathrm{P^o_{3/2}}$ ground state level.
Theoretical determination of transition rates for spin-forbidden transitions have been improved considerably over the years, but it is hard to give an exact value of the uncertainty of this rate as no measurement of the ${^4}\mathrm{D}_{5/2}$ rate for any ion in the Cl-like sequence is available (and this situation is not likely to change with the demise of beam-foil spectroscopy some years back~\citep{Traebert2008}). The estimated uncertainty of the present cited transition rate is $\leq 25\%$~\citep{Wang2020}, which will cause a maximum uncertainty of 12.5\% in the estimated magnetic field strengths. This is considerably smaller than the uncertainty introduced from the estimates of the $\Delta E$ and can at the present stage of analyses be ignored.
The M2 rate is easier to compute, but an added complication is that this decay component of the ${^4}\mathrm{D}_{7/2}$ has a different angular dependence of its polarisation pattern than the MIT (which is an electric dipole transition). One of the authors has investigated the magnetic-field-dependent angular distributions and linear polarization of E1, M2 and MIT transitions for the Ne-like ions~\citep{Li2014}. When other error sources are reduced, it should be further investigated for Fe X, but we estimate it to be negligible compared to other sources of uncertainties in the present situation.

It is clear that the proposed method is a viable candidate for direct and continuous measurements of coronal fields.
To outline a future space-based instrument designed from the proposed scheme, it would need a simple spectrometer isolating two narrow spectral regions, the short wavelength region covering the 174.543~{\AA} and 175.266~{\AA} lines and a higher wavelength region for the 257.262~{\AA} line.
The spectrometer should be intensity calibrated, similarly to the present spectrometer aboard HINODE. The other requirement that is of vital importance is the optimization of the signal to noise ratios.
As can be seen that the AR area line ratio in Fig.~\ref{fig_AR2} (b) changes from 0.72 to 0.69 over a range of magnetic field from 200 to 400~G, i.e. a factor of two change in the field strength only results in a bit over 4\% change in the line ratio. Therefore the EIS LW-SW calibration and optimization of the signal-to-noise ratio are also critical factors for the success of this technique, although at the present stage the $\Delta E$ uncertainty is clearly the most important source of error.

\section{Conclusion}
We present for the first time direct, space-based measurements of the solar corona magnetic field strength. The measurements are based on a magnetic-field induced transition, MIT, in the spectrum of Fe$^{9+}$. So far, the MIT used here is the only one known to be sensitive to the relatively weak  magnetic fields found in astrophysical plasmas such as the solar corona. The field strength we determine is around 270~G.
The most severe uncertainties come from the determination of the ${^4}\mathrm{D}$ fine structure and measured the line intensities. Both of these are possible to improve in future work, by more accurate laboratory measurement (for the former one) and an optimised design of a space-based observation (for the latter).

\acknowledgments
 This work was supported by the Swedish Research Council (VR) under Contract No. 2015-04842 and the National Natural Science Foundation of China under Contract No. 11474069. JG would like to acknowledge financial support from the the project grants ``The New Milky Way'' (2013.0052) and ``Probing charge- and mass- transfer reactions on the atomic level'' (2018.0028) from the Knut and Alice Wallenberg Foundation. We also thank J. Leenaarts for helpful discussions.

\bibliography{Fe_X}
\bibliographystyle{aasjournal}
\end{document}